%% file: main.tex
\documentclass[conference]{IEEEtran}

\usepackage{amsmath,amssymb,amsfonts,amsthm,verbatim}

\newtheorem{thm}{Theorem}
\newtheorem{lem}{Lemma}
\newtheorem{cor}{Corollary}

\begin{document}

\title{An Elementary Completeness Proof for\\Secure Two-Party Computation Primitives}

\author{\IEEEauthorblockN{Ye Wang}
\IEEEauthorblockA{Mitsubishi Electric Research Laboratories\\
Cambridge, MA, USA\\
Email: yewang@merl.com}
\and
\IEEEauthorblockN{Prakash Ishwar}
\IEEEauthorblockA{Boston University\\
Boston, MA, USA\\
Email: pi@bu.edu}
\and
\IEEEauthorblockN{Shantanu Rane}
\IEEEauthorblockA{Mitsubishi Electric Research Laboratories\\
Cambridge, MA, USA\\
Email: rane@merl.com}}

\maketitle

\begin{abstract}
In the secure two-party computation problem, two parties wish to compute a (possibly randomized) function of their inputs via an interactive protocol, while ensuring that neither party learns more than what can be inferred from only their own input and output.
For semi-honest parties and information-theoretic security guarantees, it is well-known that, if only noiseless communication is available, only a limited set of functions can be securely computed;
however, if interaction is also allowed over general communication primitives (multi-input/output channels), there are ``complete'' primitives that enable any function to be securely computed.
The general set of complete primitives was characterized recently by Maji, Prabhakaran, and Rosulek leveraging an earlier specialized characterization by Kilian.
Our contribution in this paper is a simple, self-contained, alternative derivation using elementary information-theoretic tools.
\end{abstract}

\input{intro}

\input{problem}

\input{results}

\input{common}

\input{proofs}

\bibliographystyle{IEEEtran}
\bibliography{references}

\end{document}

%% file: intro.tex
\section{Introduction}

We consider the problem of secure two-party computation, where two parties named Alice and Bob wish to correctly and privately compute outputs from their initial individual inputs, according to a (potentially randomized) function.
Correctness means that the outputs should have the appropriate conditional distribution (corresponding to the desired function) with respect to the inputs.
Privacy means that neither party should learn anything about the other party's input and output besides what can be inferred from only their own input and output.
The aim is to construct an interactive protocol that computes the desired function while satisfying these security goals.
We restrict our attention to passive (``honest but curious'') parties who will faithfully execute a given protocol, but attempt to extract additional information from their views of the execution.
However, we require information-theoretic privacy, providing unconditional security guarantees against adversaries with even unbounded computational power.

We focus on the {\em feasibility} of constructing protocols for general secure computation when the parties are allowed unlimited interaction via noise-free communication as well as via a given set of communication {\em primitives}\footnote{Primitives and functions are the same class of mathematical objects (random channels where each party has an input and output), but we use ``primitives'' to refer to the channels available for implementing a protocol, while ``function'' refers to the secure computation objective of the protocol.}, which are general memoryless two-way channels where each party may have an input and an output.
In the ``from scratch'' scenario, where only noise-free communication is allowed and no additional primitives are available, it is well-known that not all functions can be securely computed by two parties (see~\cite{BenOrGwW-ACM88-CTNCFTDC} for example).
However, given the availability of certain {\em complete} primitives, protocols can be constructed to perform any general computation.
Oblivious transfer\footnote{Oblivious transfer is the channel where Alice has a two-bit input and no output, and Bob's binary input selects one of Alice's bits to be his output.} is a complete primitive~\cite{Kilian-ACM88-CryptoFromOT}, as is any primitive that enables secure computation of oblivious transfer as the desired function~\cite{CrepeauK-FOCS-88-OTusingWeakSecure}.
Identifying complete primitives (and proposing efficient constructions) has been an active area of research with several works characterizing the complete primitives within specific subclasses:
one-way channels (primitives with one input and one output)~\cite{Crepeau-EuroCrypt-97-CrypBasedOnNoisyChans,CrepeauMW-SCN04-OTfromAnyChan,NascimentoW-IT2008-OTCapOfNoisyRes},
joint sources (primitives with no inputs)~\cite{NascimentoW-IT2008-OTCapOfNoisyRes},
and primitives with only one output or a common output~\cite{Kilian-STOC00-CryptoFromOT}.
Recently, a general characterization of all complete primitives for the passive secure two-party computation problem was given in~\cite{MajiPR-IndoCrypt12-Completeness} by leveraging the specialized results of~\cite{Kilian-STOC00-CryptoFromOT}.

Our main contribution is a simple, self-contained, alternative derivation of the general characterization of complete primitives using elementary information-theoretic tools, which contrast with the detailed combinatorial analysis of protocol structure given by~\cite{Kilian-STOC00-CryptoFromOT} and leveraged by~\cite{MajiPR-IndoCrypt12-Completeness}.
Our converse proof is based on considering the subclass of secure two-party {\em sampling} problems, where Alice and Bob have no initial inputs and only wish to generate outputs according to a desired joint distribution.
This proof technique also clarifies a subtlety: that a set of primitives that are each individually incomplete cannot provide completeness when available together.
Further, we observe that the secure sampling problems exhibit a zero-one law, in the sense that
any set of primitives is either complete or ``useless'', i.e., allowing only a set of ``trivial'' distributions to be sampled.
The trivial distributions are those that can be securely sampled from scratch, and were characterized in~\cite{WangIshwarISIT11} as those for which the mutual information is equal to the common information\footnote{This property is equivalent for the Wyner~\cite{Wyner-75-CommonInfo} and G\'acs-K\"orner~\cite{GacsKorn-73-CommonInfo} notions of common information.}.

%% file: problem.tex
\section{Problem Formulation}

\subsection{Secure Two-Party Computation Protocols}

Alice and Bob respectively start with inputs $Q$ and
$T$ with joint distribution $P_{Q,T}$ over the finite alphabet
$\mathcal{Q} \times \mathcal{T}$.
They wish to securely compute the (in general randomized) function $P_{X,Y|Q,T}$.
To realize this goal, they execute a two-party computation protocol at
the end of which Alice outputs $\hat{X} \in \mathcal{X}$ and Bob
outputs $\hat{Y} \in \mathcal{Y}$.

A protocol may involve multiple rounds of local computation
interspersed with rounds of interaction via error-free communication
or through one of the available communication primitives.
A {\em communication primitive} is a channel with input $(A,B)$ in
the finite alphabet $\mathcal{A} \times \mathcal{B}$, output $(U,V)$
in the finite alphabet $\mathcal{U} \times \mathcal{V}$, and a
conditional distribution $P_{U,V|A,B}$.  Each primitive usage is
``memoryless'', and Alice controls input $A$ and receives output $U$,
while Bob controls input $B$ and receives output $V$.
After the protocol terminates, Alice and Bob generate their respective
outputs via deterministic functions of their respective \emph{views} of
the protocol. A party's view consists of its initial input, local computations,
messages sent/received, and inputs/outputs to/from the used
primitives.

For simplicity, we only consider protocols that terminate in a fixed
(deterministic) number of rounds $n$, but do not put a bound on $n$.
A protocol consists of a sequence of steps that governs how the views
of the parties can evolve during the protocol's execution. The initial
views of Alice and Bob are their respective inputs and denoted by $(R_0,S_0) := (Q,T)$.
Let $(R_1,S_1), \ldots, (R_n, S_n)$ denote the sequence of their evolving views over $n$ rounds.
In each round $t$ of the
protocol, the evolution of views from $(R_{t-1}, S_{t-1})$ to $(R_t,
S_t)$ occurs via one of three possible structured mechanisms: local
computation, error-free message passing, or primitive usage (if
available).
\begin{itemize}
\item (Local computation) $R_t = (R_{t-1}, A)$ and $S_t = (S_{t-1},
  B)$, where $A \leftrightarrow R_{t-1} \leftrightarrow S_{t-1}
  \leftrightarrow B$ is a Markov chain.
\item (Message passing) $R_t = (R_{t-1}, g(S_{t-1}))$ and $S_t =
  (S_{t-1}, f(R_{t-1}))$, where $f$ and $g$ are some deterministic
  functions.
\item (Primitive usage) $R_t = (R_{t-1}, U)$ and $S_t = (S_{t-1}, V)$,
  where $(U,V)$ are the outputs of one of the given communication
  primitives, with inputs $A = f(R_{t-1})$ and $B = g(S_{t-1})$
  generated via some deterministic functions $f$ and $g$, and $P_{U,V|A,B}$
  corresponds to the distribution governing the primitive used.  The
  memoryless behavior of the primitives implies that $(U,V)
  \leftrightarrow (A,B) \leftrightarrow (R_{t-1}, S_{t-1})$ is a
  Markov chain.
\end{itemize}
After $n$ rounds, outputs are generated deterministically from the
final views, that is, $\hat{X} = \phi(R_n)$ and $\hat{Y} = \psi(S_n)$,
for some functions $\phi$ and $\psi$.

\subsection{Security Definitions}

A protocol for computing $P_{X,Y|Q,T}$ is called $\epsilon$-\emph{correct} if and only if the following maximal variational
distance does not exceed $\epsilon$:
\begin{align*}
\max_{P_{Q,T}} d(P_{\hat{X},\hat{Y} | Q,T} P_{Q,T}, P_{X,Y|Q,T} P_{Q,T}) \leq \epsilon,
\end{align*}
where the variational distance is given by $d(P_{\hat{Z}},P_Z) := \frac{1}{2} \sum_z | P_{\hat{Z}}(z) - P_Z(z) |$.
A protocol is $\delta$-\emph{private} if and only if the maximal information
leakage of the final views satisfies
\[
\max_{P_{Q,T}} I(R_n ; \hat{Y}, T | \hat{X}, Q) + I(S_n ; \hat{X}, Q | \hat{Y}, T) \leq \delta.
\]
We will say that a protocol is $(\epsilon,\delta)$-secure if and only if it is
$\epsilon$-correct and $\delta$-private.  
A function $P_{X,Y|Q,T}$ is said to be \emph{securely computable} given a set of primitives if and only if for all $\epsilon,\delta > 0$, there exists a protocol for computing $P_{X,Y|Q,T}$ using the given primitives that is $(\epsilon,\delta)$-secure.
A primitive is said to be \emph{complete} if and only if any function is securely computable given that primitive.
A {\em set} of primitives is said to {\em incomplete} if and only if some functions cannot be securely computed via any protocols using that set of primitives.
Note that an incomplete set must be comprised of primitives that are each individually incomplete.
The reverse implication is not immediately obvious but turns out to be true (see Theorem~\ref{thm:main}).

\subsection{Secure Two-Party Sampling}

The secure two-party {\em sampling} problem is the special case where Alice and Bob have no inputs and the goal simplifies to generating outputs with the joint distribution $P_{X,Y}$.
Their initial views are constant $R_0 = S_0 = 0$ and the conditions for $\epsilon$-correctness and $\delta$-privacy simplify to $d(P_{\hat{X},\hat{Y}}, P_{X,Y}) \leq \epsilon$ and $I(R_n ; \hat{Y} | \hat{X}) + I(S_n ; \hat{X} | \hat{Y}) \leq \delta$, respectively.

The distributions that can be securely sampled via protocols that use
only error-free communication (and no other primitives)
will be called \emph{trivial}, since they can always be securely
sampled regardless of the other primitives available.
A set of primitives is said to be \emph{useless for sampling} if only the trivial distributions can be securely sampled using that set of primitives.
Since secure sampling is a special case of secure computation, a complete primitive allows any distribution to be securely sampled.
Hence, a set of primitives is incomplete (for general computation) if it is useless for sampling.

%% file: results.tex
\section{Characterization Results}

\subsection{Preliminaries}

Common information plays a key role in the characterizations of both
the secure sampling and computation problems.  There are two related
(and somewhat complementary) notions of common information, one
introduced by Wyner~\cite{Wyner-75-CommonInfo} and the other
introduced by G\'acs-K\"orner~\cite{GacsKorn-73-CommonInfo}.  We will
review only the Wyner common information here to allow us to quickly
state our results, and leave G\'acs-K\"orner common information and
other related concepts to be reviewed later in
Section~\ref{sec:common-info}.

The Wyner common information of two random variables $(X,Y)$ is given
by
\[
C(X;Y) := \hspace{-8pt} \min_{Z: I(X;Y|Z) = 0} \hspace{-8pt} I(X,Y ; Z),
\]
where the minimum can be attained by a $Z \in \mathcal{Z}$ with
$|\mathcal{Z}| \leq |\mathcal{X} \times
\mathcal{Y}|$~\cite{Wyner-75-CommonInfo}.  This quantity characterizes
the solution of the Gray-Wyner source coding problem.  Note that, in
general, $C(X;Y) \geq I(X;Y)$~\cite{Wyner-75-CommonInfo}.

It follows from the results of \cite{WangIshwarISIT11} and the
continuity of Wyner common information (see
Lemma~\ref{lem:wyner-continuous} in Section~\ref{sec:common-info}),
that the trivial distributions, i.e., those which can be securely
sampled from scratch, are precisely those where $C(X;Y) = I(X;Y)$ (see
Lemma~\ref{lem:trivial-props} in Section~\ref{sec:common-info} for
equivalent conditions).  We will hence use the terms \emph{trivial}
(and \emph{non-trivial}) to refer to joint distributions $P_{X,Y}$
which do (and, respectively, do not) satisfy $C(X;Y) = I(X;Y)$.

\subsection{Main Results}

The main theorem characterizes the complete primitives and notes that incomplete primitives are useless for sampling.

\begin{thm} \label{thm:main}
A primitive $P_{U,V|A,B}$ is complete if and only if for uniformly distributed inputs
$(A,B) \sim \text{Unif}(\mathcal{A} \times \mathcal{B})$
$C(A,U;B,V) > I(A,U;B,V)$, where $(U,V,A,B) \sim P_{U,V|A,B} P_{A,B}$.
Further, any set of incomplete primitives is useless for sampling.
\end{thm}

An interpretation of a complete primitive is that can be used, with independent uniformly distributed inputs, to produce non-trivially distributed randomness in the resultant views $(A,U)$ and $(B,V)$.

The following corollary characterizes the feasibility of secure sampling, which exhibits a zero-one law: any set of primitives is either complete or useless for sampling.

\begin{cor}
Given any set of primitives, if at least one is complete (see
conditions in Theorem~\ref{thm:main}), then any distribution $P_{X,Y}$
can be securely sampled.  Otherwise, only the trivial distributions,
where $C(X;Y) = I(X;Y)$, can be securely sampled.
\end{cor}

%% file: common.tex
\section{Properties of Common Information} \label{sec:common-info}

This section reviews key concepts and results needed to establish our
main results. They are, however, also of independent interest.

The {\em graphical representation} of $P_{X,Y}$ is the bipartite graph
with an edge between $x \in \mathcal{X}$ and $y \in \mathcal{Y}$ if
and only if $P_{X,Y}(x,y) > 0$.  The {\em common part} of two random
variables $(X,Y)$, denoted by $W_{X,Y}$, is the (unique) label of the
connected component of the graphical representation of $P_{X,Y}$ in
which $(X,Y)$ falls. Note that $W_{X,Y}$ is a deterministic function
of $X$ alone and also a deterministic function of $Y$ alone.

The G\'acs-K\"orner common information of two random variables $(X,Y)$
is given by $K(X;Y) := H(W_{X,Y})$~\cite{GacsKorn-73-CommonInfo}.  The
operational significance of $K(X;Y)$ is that it is the maximum number
of common bits per symbol that can be independently extracted from $X$
and $Y$.  Note that, in general, $K(X;Y) \leq
I(X;Y)$~\cite{GacsKorn-73-CommonInfo}.

While it may be tedious, in general, to solve the optimization problem
that defines Wyner common information, one can conveniently check if
it is equal to its lower bound by using its well-known relationship to
G\'acs-K\"orner common information and other properties given in the
following lemma (see~\cite{AhlsKorn-74-CommonInfo}).

\begin{lem} {\rm \cite{AhlsKorn-74-CommonInfo}} \label{lem:trivial-props}
For any random variables $(X,Y)$, the following are equivalent:
\begin{enumerate}
\item $C(X;Y) = I(X;Y)$,
\item $K(X;Y) = I(X;Y)$,
\item There exists $Z$ such that $Z \leftrightarrow X \leftrightarrow
  Y$, $Z \leftrightarrow Y \leftrightarrow X$, and $X \leftrightarrow
  Z \leftrightarrow Y$ are all Markov chains,
\item $X \leftrightarrow W_{X,Y} \leftrightarrow Y$ is a Markov chain,
  where $W_{X,Y}$ is the common part of $(X,Y)$.
\end{enumerate}
\end{lem}

One can also determine whether common information is equal to mutual
information by checking if conditional entropy is positive after
``removing redundancies'' from the random variables.  To \emph{remove
  redundancy} from $X$ with respect to $P_{X,Y}$, first partition the
support of $P_{X}$ into equivalence classes using $P_{Y|X=x} =
P_{Y|X=x'}$ as the equivalence rule for $x, x' \in \mathcal{X}$, then
uniquely label these classes and define $\tilde{X}$ as the label of
the class in which $X$ falls.  Similarly, $\tilde{Y}$ can be defined
as $Y$ with redundancies removed.  Note that, by construction, $X
\leftrightarrow \tilde{X} \leftrightarrow \tilde{Y} \leftrightarrow Y$
is a Markov chain.

\begin{lem} \label{lem:redundancy}
For any random variables $(X,Y)$, the following are equivalent:
\begin{enumerate}
\item $C(X;Y) = I(X;Y) = K(X;Y)$, \label{cond:trivial}
\item $H(\tilde{X} | \tilde{Y}) = 0$, \label{cond:XgivenY}
\item $H(\tilde{Y} | \tilde{X}) = 0$, \label{cond:YgivenX}
\end{enumerate}
where $(\tilde{X},\tilde{Y})$ are $(X,Y)$ with redundancies removed.
\end{lem}

\begin{IEEEproof}
This lemma can be shown to follow from the monotone region results of~\cite{PrabX2-TransIT-AsstComInfo}. We, however, provide a simpler, self-contained proof here.
Any $x,x' \in \mathcal{X}$ with $P_{Y|X=x} = P_{Y|X=x'}$ are clearly
in the same connected component of the graphical representation of
$P_{X,Y}$.  If $X \leftrightarrow W_{X,Y} \leftrightarrow Y$ is a
Markov chain, then for any symbols $x,x' \in \mathcal{X}$ attached to
the same connected component, $P_{Y|X=x} = P_{Y|X=x'}$.  Thus, given
condition~\ref{cond:trivial}, we find that $W_{X,Y}$, $\tilde{X}$, and
$\tilde{Y}$ (via similar arguments) are equivalent, that is, $W_{X,Y}
= f(\tilde{X}) = g(\tilde{Y})$ for some bijective functions $f$ and
$g$.  Hence, it follows that condition~\ref{cond:trivial} implies
condition~\ref{cond:XgivenY} and~\ref{cond:YgivenX}.  Given
condition~\ref{cond:XgivenY}, $\tilde{X}$ is a function of
$\tilde{Y}$, and hence a function of $Y$. By construction, $\tilde{X}$ is a function of $X$, and $X
\leftrightarrow \tilde{X} \leftrightarrow \tilde{Y} \leftrightarrow Y$
is a Markov chain. Hence, $X \leftrightarrow \tilde{X} \leftrightarrow
Y$, $\tilde{X} \leftrightarrow X \leftrightarrow Y$, and $\tilde{X}
\leftrightarrow Y \leftrightarrow X$ are all Markov chains and
condition~\ref{cond:trivial} holds by Lemma~\ref{lem:trivial-props}.
Similarly, condition~\ref{cond:YgivenX} also implies
condition~\ref{cond:trivial}. 
\end{IEEEproof}

Another useful property for checking whether the Wyner common
information is close to the mutual information is given in the next
lemma from~\cite{WangIshwarISIT11}.

\begin{lem} {\rm \cite{WangIshwarISIT11}} \label{lem:almost-trivial}
For any random variables $(X,Y)$, $C(X;Y) - I(X;Y) \leq \delta$ if and
only if there exist $Z$ such that $X \leftrightarrow Z \leftrightarrow
Y$ is a Markov chain, and $I(Z;X|Y) + I(Z;Y|X) \leq \delta$.
\end{lem}

Wyner common information is a uniformly continuous functional of
$P_{X,Y}$ for all $P_{X,Y}$ as established in the next lemma.
The G\'acs-K\"orner common information, in contrast, is discontinuous.

\begin{lem} \label{lem:wyner-continuous}
If $P_{X,Y}, P_{\hat{X},\hat{Y}}$ are joint distributions over the
same finite alphabet $\mathcal{X} \times \mathcal{Y}$ with variational
distance $d(P_{\hat{X},\hat{Y}}, P_{X,Y}) \leq \epsilon$, then $|C(X;Y)
- C(\hat{X};\hat{Y})| \leq \alpha(\epsilon)$, for some function
$\alpha$ where $\alpha(\epsilon) \longrightarrow 0$ as $\epsilon
\longrightarrow 0$.
\end{lem}

\begin{IEEEproof}
One can construct random variables $(X,Y) \sim P_{X,Y}$ and
$(\hat{X},\hat{Y}) \sim P_{\hat{X},\hat{Y}}$ such that
$\Pr\big((\hat{X},\hat{Y}) \neq (X,Y)\big) = d(P_{\hat{X},\hat{Y}},
P_{X,Y})$~\cite{Zhang-IT07}.  Let $Z$ be the random
variable such that $C(X;Y) = I(X,Y ; Z)$ and $X \leftrightarrow Z
\leftrightarrow Y$ is a Markov chain.  Next, let
\[
\hat{Z} := \begin{cases}
(Z, \perp, \perp), & \text{when }(\hat{X},\hat{Y}) = (X,Y),\\
(\perp, \hat{X}, \hat{Y}), & \text{when }(\hat{X},\hat{Y}) \neq (X,Y),
\end{cases}
\]
where $\perp$ is a constant symbol not in the alphabets $\mathcal{X}$,
$\mathcal{Y}$, or $\mathcal{Z}$.  By construction, $\hat{X}
\leftrightarrow \hat{Z} \leftrightarrow \hat{Y}$ is a Markov chain,
and $\Pr\big( (\hat{X},\hat{Y},\hat{Z}) \neq (X,Y,(Z,\perp,\perp))
\big) \leq \epsilon$.  Thus,
\begin{align*}
C(\hat{X};\hat{Y}) \leq I(\hat{X}, \hat{Y};\hat{Z}) &\leq I(X, Y; Z) + \alpha(\epsilon) \\
&= C(X;Y) + \alpha(\epsilon)
\end{align*}
for some $\alpha(\epsilon)$ with $\alpha(\epsilon) \longrightarrow 0$
as $\epsilon \longrightarrow 0$, where the second inequality follows
due to the uniform continuity of entropy~\cite{Zhang-IT07}.
Symmetrically, we can argue that $C(X;Y) \leq C(\hat{X};\hat{Y}) +
\alpha(\epsilon)$, and hence $|C(X;Y) - C(\hat{X};\hat{Y})| \leq
\alpha(\epsilon)$.
\end{IEEEproof}

%% file: proofs.tex
\section{Proof of Theorem~\ref{thm:main}}

\subsection{Converse Result}

We will show that, given any set of primitives that each fail
to satisfy the completeness conditions, only trivial distributions can
be securely sampled, and hence the primitives are incomplete and
useless.
The first part of our converse proof is closely related to the method
of monotones -- functionals that are monotonic over the sequence of
views -- introduced in~\cite{WolfWull-08-Monotones}. Specifically, we
will show that the distributions of the views $P_{R_t,S_t}$ will
remain trivial throughout the execution of the protocol. Then, we will
argue that given final views $(R_n, S_n)$ with a trivial distribution,
only ``almost trivial'' (in the sense of Wyner common information
being close to mutual information) outputs can be securely produced by
a $\delta$-private protocol.  This result, in conjunction with the
continuity of Wyner common information
(see Lemma~\ref{lem:wyner-continuous}), implies that only trivial
distributions can be securely sampled.

The next two lemmas establish that if we start with views $(R_{t-1},
S_{t-1})$ that have a trivial distribution, then the views $(R_t,
S_t)$, after respectively local computation and message passing, must
also have a trivial distribution.
These two lemmas can be shown to follow from results in~\cite{WolfWull-08-Monotones}; however, we give short, self-contained proofs here.

\begin{lem} \label{lem:local-comp}
Let $C(R;S) = I(R;S)$. If $A \leftrightarrow R \leftrightarrow S
\leftrightarrow B$ is a Markov chain then $C(A,R;B,S) = I(A,R;B,S)$.
\end{lem}

\begin{IEEEproof}
Let $W_{R,S}$ be the common part of random variables $(R,S)$. Since
$C(R;S) = I(R;S)$, it follows that $R \leftrightarrow W_{R,S}
\leftrightarrow S$ is a Markov chain.  Since $W_{R,S}$ is a function
of $R$ alone and $S$ alone, it also follows that $(A,R)
\leftrightarrow W_{R,S} \leftrightarrow (B,S)$ is a Markov
chain. Hence $C(A,R;B,S) = I(A,R;B,S)$ by
Lemma~\ref{lem:trivial-props}.
\end{IEEEproof}

\begin{lem} \label{lem:msg-pass}
Let $C(R;S) = I(R;S)$. If $f,g$ are deterministic functions then
$C(R,g(S);S,f(R)) = I(R,g(S);S,f(R))$.
\end{lem}

\begin{IEEEproof}
Let $W_{R,S}$ be the common part of $(R,S)$ and $Z := (W_{R,S}, f(R),
g(S))$.  Since $Z$ is a function of $(R,g(S))$ alone and $(S,f(R))$
alone, $(R,g(S)) \leftrightarrow (S,f(R)) \leftrightarrow Z$ and
$(S,f(R)) \leftrightarrow (R,g(S)) \leftrightarrow Z$ are both Markov
chains. Since
\begin{align*}
&I(R, g(S); S, f(R) | Z) = I(R; S | W_{R,S}, f(R), g(S)) \\
&\quad \leq I(R,f(R); S, g(S) | W_{R,S}) = I(R; S | W_{R,S}) = 0,
\end{align*}
it follows that $(R,g(S)) \leftrightarrow Z \leftrightarrow (S,f(R))$
is a Markov chain. Hence, $C(R,g(S);S,f(R)) = I(R,g(S);S,f(R))$ by
Lemma~\ref{lem:trivial-props}.
\end{IEEEproof}

The following Lemma~\ref{lem:simplification} establishes that if we start from views $(R_{t-1}, S_{t-1})$ with a
trivial distribution, then using a primitive that does not meet the completeness
conditions, with any inputs $A = f(R_{t-1})$ and $B = g(S_{t-1})$, results in views
$(R_t, S_t) := ((R_{t-1}, U), (S_{t-1}, V))$ that also have a trivial distribution.
First, we show an auxiliary result in Lemma~\ref{lem:simp_aux} to facilitate the proof of Lemma~\ref{lem:simplification}.

\begin{lem} \label{lem:simp_aux}
If a primitive $P_{U,V|A,B}$ does not meet the completeness conditions of
Theorem~\ref{thm:main}, then for all random variables $A \in \mathcal{A}$, $B \in \mathcal{B}$, $Z$ such that $A \leftrightarrow Z \leftrightarrow B$ forms a Markov chain, we have that $C(Z,A,U;Z,B,V) = I(Z,A,U;Z,B,V)$, where $(U,V,A,B,Z) \sim P_{U,V|A,B}P_{A,B,Z}$.
\end{lem}

\begin{IEEEproof}
To facilitate this proof, we first elaborate upon some general properties of the graphical representation of a trivial distribution $P_{X,Y}$ where $C(X;Y) = I(X;Y)$.
The {\em weighted} graphical representation is the bipartite graph that has an edge with weight $P_{X,Y}(x,y)$ between every $x \in \mathcal{X}$ and $y \in \mathcal{Y}$ where $P_{X,Y}(x,y) > 0$.
Recall that the common part $W_{X,Y}$ is the unique label of the connected component of the graphical representation in which $(X,Y)$ occurs, and can be expressed as $W_{X,Y} = \phi(X) = \psi(Y)$ for deterministic functions $\phi$ and $\psi$ that indicate the connected component in which $X$ and $Y$, respectively, occur.
Lemma~\ref{lem:trivial-props} establishes that triviality is equivalent with $X \leftrightarrow W_{X,Y} \leftrightarrow Y$.
This Markov chain is equivalent to the property that within each connected component of the weighted graphical representation of $P_{X,Y}$, each edge weight $P_{X,Y}(x,y)$ in that component can be factored as the product of a function of $x$ and a function of $y$, namely, $P_{X,Y}(x,y) = P_{W} (\phi(x)) P_{X|W} (x | \phi(x) ) P_{Y|W} (y | \psi(y))$, for $(x,y)$ such that $\phi(x) = \phi(y)$.
Thus, we can show that a distribution is trivial by determining that, within each connected component of its graphical representation, the edge weights can be factored.

Let $(\hat{U},\hat{V},\hat{A},\hat{B}) \sim P_{U,V|A,B} P_{\hat{A},\hat{B}}$, where $P_{\hat{A},\hat{B}} = P_{\hat{A}} P_{\hat{B}}$ is the uniform distribution over $\mathcal{A} \times \mathcal{B}$, that is $(\hat{A}, \hat{B})$ are independent uniformly distributed inputs and $(\hat{U}, \hat{V})$ are the corresponding outputs generated by the primitive.
Consider the weighted graphical representation of the joint distribution between $(\hat{A},\hat{U})$ and $(\hat{B},\hat{V})$.
Since the primitive does not meet the completeness conditions, we have that the edge weights within each connect component can be factored, that is,
\begin{align*}
& P_{(\hat{A},\hat{U}),(\hat{B},\hat{V})} ((a,u),(b,v)) =\\
& \quad P_{\hat{W}} (\hat{\phi}(a,u)) P_{\hat{A},\hat{U}|\hat{W}}(a,u|\hat{\phi}(a,u)) P_{\hat{B},\hat{V}|\hat{W}} (b,v|\hat{\psi}(b,v)),
\end{align*}
for $(a,b,u,v)$ such that $\hat{\phi}(a,u) = \hat{\psi}(b,v)$,
where $(\hat{W},\hat{\phi},\hat{\psi})$ denotes the common part and its corresponding indicator functions.
Since $P_{(\hat{A},\hat{U}),(\hat{B},\hat{V})} = P_{U,V|A,B} P_{\hat{A}} P_{\hat{B}}$, we have that $P_{U,V|A,B}$ can also similarly be factored for the edges within each connected component.

Now consider any independent inputs $(\tilde{A}, \tilde{B}) \sim P_{\tilde{A}} P_{\tilde{B}}$ and the corresponding primitive outputs $(\tilde{U}, \tilde{V})$, that is, $(\tilde{U}, \tilde{V}, \tilde{A}, \tilde{B}) \sim P_{U,V|A,B} P_{\tilde{A}} P_{\tilde{B}}$.
Note that the connectivity of the graphical representation of the joint distribution between $(\tilde{A},\tilde{U})$ and $(\tilde{B},\tilde{V})$ is strongly related to the connectivity of that between $(\hat{A},\hat{U})$ and $(\hat{B},\hat{V})$.
Specifically, if $P_{\tilde{A}}$ and $P_{\tilde{B}}$ are both full support, the connectivity is exactly the same, and otherwise, only some edges are removed for the values $(a,b)$ that are not in the support.
Thus, for each connected component in the graphical representation of $P_{((\tilde{A},\tilde{U}), (\tilde{B},\tilde{V}))}$, it edges are fully contained within a single connected component in the graphical representation of $P_{((\hat{A},\hat{U}), (\hat{B},\hat{V}))}$.
Hence, within each connected component, the edge weights, which are given by $P_{((\tilde{A},\tilde{U}), (\tilde{B},\tilde{V}))} = P_{U,V|A,B} P_{\tilde{A}} P_{\tilde{B}}$, also can be factored for the edges within each connected component, due to the factorability of $P_{U,V|A,B}$ established above.

Finally, let $(A, B, Z)$ be any random variables such that $A \leftrightarrow Z \leftrightarrow B$ forms a Markov chain and $(U,V)$ be the corresponding primitive outputs, where $(U,V,A,B,Z) \sim P_{U,V|A,B} P_{A,B,Z}$.
Consider the graphical representation of the joint distribution between $(Z,A,U)$ and $(Z,B,V)$.
Note that $Z$ appears on both sides, which partitions the graphical representation into disjoint subgraphs for each $z$ where $P_{Z}(z) > 0$.
For each such subgraph, the edge weights are given by $P_{U,V|A,B} P_{A|Z = z} P_{B | Z = z} P_{Z}(z)$.
Thus, each subgraph is isomorphic to the weighted graphical representation for independent inputs $(\tilde{A},\tilde{B}) \sim P_{A|Z = z} P_{B | Z = z}$, except with the edges scaled by $P_{Z}(z)$.
Due to the factorability for independent inputs established above, we have that within each connected component within each subgraph, the edge weights can be factored.
Since each connected component of the overall graphical representation must clearly be contained within one of the disjoint subgraphs, we have that the edge factorability property holds within all connected components of the weighted graphical representation, and hence we have that $C(Z,A,U;Z,B,V) = I(Z,A,U;Z,B,V)$.
\end{IEEEproof}

\begin{lem} \label{lem:simplification}
If a primitive $P_{U,V|A,B}$ does not meet the completeness conditions of
Theorem~\ref{thm:main}, then for all random variables $(R,S)$ such that $C(R;S) = I(R;S)$ and functions $f : \mathcal{R} \rightarrow \mathcal{A}$, $g : \mathcal{S} \rightarrow \mathcal{B}$, we have that $C(R,U;S,V) = I(R,U;S,V)$, where $A = f(R)$, $B = g(S)$, and $(U,V,A,B) \sim P_{U,V|A,B}P_{A,B}$.
\end{lem}

\begin{IEEEproof}
Let $W_{R,S}$ be the common part of $(R,S)$. By Lemma~\ref{lem:trivial-props}, we have that $(A,R) \leftrightarrow W_{R,S} \leftrightarrow (B,S)$ is a Markov chain.
Thus, by Lemma~\ref{lem:simp_aux}, 
we have that $C(W_{R,S}, A, U; W_{R,S}, B, V) = I(W_{R,S}, A, U; W_{R,S}, B, V)$.
Since $(U,V) \leftrightarrow (A,B) \leftrightarrow (R,S,W_{R,S})$ is a Markov chain,  $R \leftrightarrow (W_{R,S},A,U) \leftrightarrow (W_{R,S},B,V) \leftrightarrow S$ is also a Markov chain.
Thus, by Lemma~\ref{lem:local-comp}, we have that $C(R,U;S,V) = C(R, W_{R,S}, A, U; S, W_{R,S}, B, V) = I(R, W_{R,S}, A, U; S, W_{R,S}, B, V) = I(R,U;S,V)$.
\end{IEEEproof}

Combining Lemmas~\ref{lem:local-comp},~\ref{lem:msg-pass}, and~\ref{lem:simplification}, and noting that
the initial views $(R_0, S_0) := (0,0)$ are trivial, we can conclude that the
final views $(R_n, S_n)$ also have a trivial distribution.

The next lemma establishes that for any $\delta$-private protocol, if the
final views have a trivial distribution, then the outputs must satisfy
$C(\hat{X};\hat{Y}) - I(\hat{X};\hat{Y}) \leq \delta$.

\begin{lem} \label{lem:secure-extraction}
Let $C(R;S) = I(R;S)$. If $(\phi,\psi)$ are deterministic functions
such that $I(R; \psi(S) | \phi(R)) + I(S; \phi(R) | \psi(S)) \leq
\delta$ then $C(\phi(R); \psi(S)) - I(\phi(R); \psi(S)) \leq \delta$.
\end{lem}

\begin{IEEEproof}
Let $W_{R,S}$ be the common part of $(R,S)$. Since $\phi$ and $\psi$
are deterministic functions, it follows that $\phi(R) \leftrightarrow
W_{R,S} \leftrightarrow \psi(S)$ is a Markov chain.  Using the
property that $W_{R,S}$ is a function of $R$,
\begin{align*}
&I(W_{R,S} ; \psi(S) | \phi(R)) \\
&\quad = H(\psi(S) | \phi(R)) - H(\psi(S) | \phi(R), W_{R,S}) \\
&\quad \leq H(\psi(S) | \phi(R)) - H(\psi(S) | \phi(R), R) \\
&\quad = I(R; \psi(S) | \phi(R)).
\end{align*}
Similarly, $I(W_{R,S} ; \phi(R) | \psi(S)) \leq I(S; \phi(R) | \psi(S))$ follows.
Thus, $I(W_{R,S} ; \psi(S) | \phi(R)) + I(W_{R,S} ; \phi(R) | \psi(S))
\leq \delta$, and hence, $C(\phi(R); \psi(S)) - I(\phi(R); \psi(S))
\leq \delta$ by Lemma~\ref{lem:almost-trivial}.
\end{IEEEproof}

Thus, if $P_{X,Y}$ can be securely sampled given a set of primitives
that do not satisfy the completeness conditions, then for any
$\epsilon, \delta > 0$ there exists $P_{\hat{X},\hat{Y}}$ such that
$d(P_{\hat{X},\hat{Y}}, P_{X,Y}) \leq \epsilon$ and
$C(\hat{X};\hat{Y}) - I(\hat{X};\hat{Y}) \leq \delta$.  Finally, due
to the continuity of Wyner common information (see
Lemma~\ref{lem:wyner-continuous}) and entropy, it follows that
$P_{X,Y}$ must be trivial.

\subsection{Achievability Sketch} \label{sec:achievability}

Due to space restrictions and since the essential techniques are
well-known in the literature, we will only sketch the overall scheme
for securely computing any function given a primitive satisfying
the completeness conditions.  Also, we aim only to describe a general
but straight-forward approach to show feasibility.  Of course, more
complex approaches or specialized methods exploiting the structure of
particular problem instances may yield more efficient schemes.  The
overall achievability argument follows these high-level steps:
\begin{enumerate}
\item Given a primitive satisfying the completeness conditions, we can
  construct a protocol which can simulate a source primitive $P_{U,V}$
  that has a non-trivial distribution.\label{step:use-primitive}
\item The simulated source primitive with a non-trivial distribution
  can be converted into a binary erasure source via the methods
  of~\cite{NascimentoW-IT2008-OTCapOfNoisyRes}.\label{step:BES-conversion}
\item Continuing with the methods
  of~\cite{NascimentoW-IT2008-OTCapOfNoisyRes}, the binary erasure
  source can be used to perform oblivious
  transfers.\label{step:perform-OT}
\item Using the methods of~\cite{Kilian-ACM88-CryptoFromOT}, general
  secure computation can be performed via the oblivious transfers.\label{step:sample-from-OT}
\end{enumerate}
We further explain these steps below.

\emph{Step~\ref{step:use-primitive})} Given a primitive satisfying the completeness conditions, Alice and Bob can use the primitive to simulate a source primitive (one with no inputs) with a non-trivial distribution by respectively generating independent, uniform inputs $A \sim \text{Unif}(\mathcal{A})$ and $B \sim \text{Unif}(\mathcal{B})$.
This procedure results in Alice and Bob respectively holding $\hat{U} := (A,U)$
and $\hat{V} := (B,V)$ that have the non-trivial distribution $P_{\hat{U},\hat{V}} := P_{(A,U),(B,V)}$, and can be independently repeated to generate an iid sequence of sample pairs from the non-trivial distribution $P_{\hat{U},\hat{V}}$.

\emph{Step~\ref{step:BES-conversion})} The methods
of~\cite{NascimentoW-IT2008-OTCapOfNoisyRes} require a source
primitive $P_{U,V}$ with $H(\tilde{U} | \tilde{V}) > 0$ where
$(\tilde{U},\tilde{V})$ are the random variables $(U,V)$ with
redundancies removed.  However, by Lemma~\ref{lem:redundancy}, this is
equivalent to requiring a source primitive with a non-trivial
distribution.  Due to the properties of distributions with
$H(\tilde{U} | \tilde{V}) > 0$, sample pairs from this non-trivial
source can be selectively discarded, leaving behind sample pairs that
essentially have a binary erasure source distribution, where Alice's
sample is a uniform bit and Bob's sample is either equal to Alice's or
an erasure symbol (see~\cite{NascimentoW-IT2008-OTCapOfNoisyRes} for
details).

\emph{Step~\ref{step:perform-OT})} Using these binary erasure source
sample pairs, one can perform oblivious transfer, that is, to
essentially simulate the primitive $P_{U,V|A,B}$ where $A := (A_0,
A_1)$, $A_0,A_1,B \in \{0,1\}$, and $(U,V) := (0, A_B)$
(see~\cite{NascimentoW-IT2008-OTCapOfNoisyRes}). Bob first chooses two sample
pairs of the binary erasure source for which there is exactly one
erasure, and then instructs Alice to respectively exclusive-or her two
input bits $(A_0, A_1)$ with the two corresponding bits she has from
her half of the erasure source such that the non-erased bit is aligned
with the input that Bob wants (according to $B$).  By sending the
result to Bob over the error-free channel, he can recover $A_B$, while
Alice's other bit is masked due to the erasure.

\emph{Step~\ref{step:sample-from-OT})} Using the methods
of~\cite{Kilian-ACM88-CryptoFromOT}, the ability to perform oblivious
transfers can be leveraged to compute any secure computation.
For approximating $P_{X,Y|Q,T} P_{Q,T}$ within
any variational distance $\epsilon > 0$, the outputs
$(\hat{X},\hat{Y})$ could be computed from a boolean circuit with a
uniformly random sequence of bits as input.  Each party first
independently generates a uniformly random sequence of bits.  Using
these as shares of the input sequence, the parties then apply the
methods of~\cite{Kilian-ACM88-CryptoFromOT} for securely evaluating
the circuit to generate their respective outputs.

Note that evaluating the circuit in the last step requires a fixed
number of oblivious transfers; however, the number that can actually
be performed depends on the random number of binary erasure sample
pairs extracted in the second step.  With a protocol of fixed length
(and hence fixed primitive usages), the situation of insufficient
erasure samples can be handled as an error event leading to a constant
output, and its effect can be made asymptotically small and hence
within any $\epsilon$ approximation error.  This approach also has the
benefit of yielding constructions that are perfectly private ($\delta
= 0$).